\documentclass[12pt]{iopart}
\usepackage{graphicx}
\usepackage{wrapfig}

\newcommand*{\mub}{\ensuremath{\mu_{B}}}
\newcommand*{\roots}{\ensuremath{\sqrt{s_{_{NN}}}}}

\begin{document}

\title{Validity of the Hadronic Freeze-Out Curve}

\author{F~Becattini,$^{1}$ M~Bleicher,$^{2}$ T~Kollegger,$^{2}$ M~Mitrovski,$^{3}$ T~Schuster,$^{2,4}$ and R~Stock$^{2,4,\ast}$}

\address{$^{1}$ Universit\'a di Firenze and INFN Sezione di Firenze}
\address{$^{2}$ Frankfurt Institute for Advanced Studies(FIAS)}
\address{$^{3}$ Physics Dept., BNL, Brookhaven}
\address{$^{4}$ Institut fuer Kernphysik, University of Frankfurt}

\ead{$^{\ast}$ stock@ikf.uni-frankfurt.de}

\begin{abstract}
We analyze hadro-chemical freeze-out in central Pb+Pb collisions at CERN SPS energies, employing the hybrid version of UrQMD which models hadronization by the Cooper-Frye mechanism, and matches to a final hadron-resonance cascade. We fit the results both before and after the cascade stage using the Statistical Hadronization Model, to assess the effect of the cascade phase. We observe a strong effect on antibaryon yields except anti-$\Omega$, resulting in a shift in $T$ and \mub. We discuss the implications for the freeze-out curve.
\end{abstract}


Hadron production in relativistic A+A collisions is supposed, since Bevalac times~\cite{Montvay:1978sv,Stock:1985xe}, to proceed via two separate freeze-out stages. The first, ``hadro-chemical'' freeze-out fixes the hadronic yields per species, and their ratios, which are conserved throughout the subsequent hadron-resonance cascade expansion. At its end, ``kinetic freeze-out'' delivers the eventually observed bulk properties such as $p_{\mathrm{T}}$ spectra, HBT correlations, collective flow properties, etc.. Most remarkably, the hadronic yield distributions over species is understood to resemble a grand canonical statistical Gibbs equilibrium ensemble~\cite{Becattini:2003wp,PBM_QGP3}, from AGS up to RHIC/LHC energies.
Its two most relevant parameters, temperature $T$ and baryochemical potential \mub, thus capture a snapshot of the system dynamical evolution, taken at the instant of hadro-chemical freeze-out.

In relativistic A+A collisions the thus determined $T$ increases monotonically with \roots, saturating at about 170~MeV (the A+A Hagedorn temperature) while \mub\ approaches zero. Systematic statistical model (SM) analysis reveals the ``freeze-out curve''~\cite{3} in the $T,\mub$ plane, in which we usually also represent the conjectured plot of the phase diagram of QCD matter.

\begin{figure}
\begin{center}
\includegraphics[width=0.45 \textwidth]{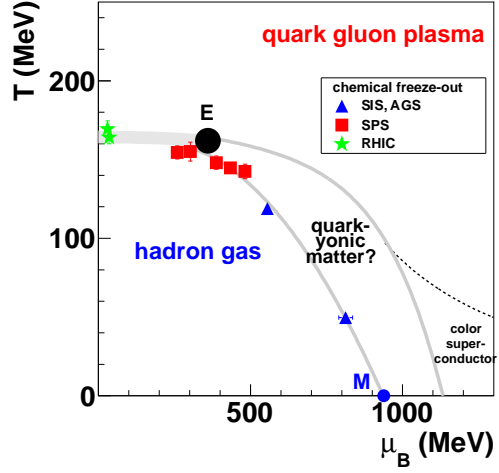}
\end{center}
\caption{Sketch of the QCD phase diagram, including the hadronic freeze-out curve.}
\label{fig:1}
\end{figure}

Such a plot is given in figure~\ref{fig:1}. It shows two principal lines, firstly a parton-hadron coexistence boundary line, inferred from lattice QCD~\cite{4} at low \mub, and from chiral restoration theory~\cite{5} at high \mub. And, second, the SM freeze-out curve. Remarkably, the lines merge toward $T=170$~MeV, $\mub=0$.
The freeze-out curve locates the QCD hadronization transition temperature $T_{\mathrm{c}}$: hadronization thus coincides with hadronic freeze-out, here. Equally remarkable, however, the two lines disentangle with increasing \mub, becoming spaced by about 30 MeV temperature difference toward $\mub=500$~MeV which corresponds to $\roots=5$~GeV in A+A collisions.

What are we freezing out from, here? A phase transition like the parton-hadron transition at small \mub\ would offer conditions that establish a grand canonical species equilibrium~\cite{6}. Recent ideas concerning a further QCD phase at high \mub, of quarkyonic matter~\cite{7},
come to mind. Indicated in figure~\ref{fig:1} is a scenario in which the freeze-out curve is identified, tentatively, with a hypothetical quarkyonic matter
phase boundary.

Before embarking on this idea a different possible situation needs to be addressed. Taking for granted that the hadron-resonance phase is indeed created at the coexistence curve it might be conceivable that an expansive hadron/resonance evolution stage, setting in at $T_{\mathrm{c}}$ and $\mu_{B,\mathrm{c}}$, cools down the population maintaining chemical equilibrium until chemical freeze-out occurs by mere dilution (the inelastic mean free path becoming longer than the system size), at lower $T$, higher \mub, thus defining the freeze-out curve.

In this note we test the latter scenario. We employ the framework of the microscopic transport model UrQMD. Its recent hybrid version~\cite{8} features a 3+1 hydrodynamic expansion during the high density stage, terminated by the Cooper-Frye hadronization mechanism once the energy density of flow cells falls below a ``critical'' energy density. This criterion resembles hadronization at the coexistence line of figure~\ref{fig:1}. The hadron/resonance population can be examined, either, by terminating the evolution at this stage, emitting into vacuum, and fitting the yield distribution by the grand canonical statistical model~\cite{9}. Alternatively, the UrQMD hadron/resonance cascade expansion stage is attached, as an ``afterburner''. The outcome is again fitted by the SM. Will the afterburner cool the system in equilibrium, to start from $T_{\mathrm{c}}$ and arrive at $T$ on the freeze-out curve?

\begin{figure}
\begin{center}
\includegraphics[width=0.4 \textwidth]{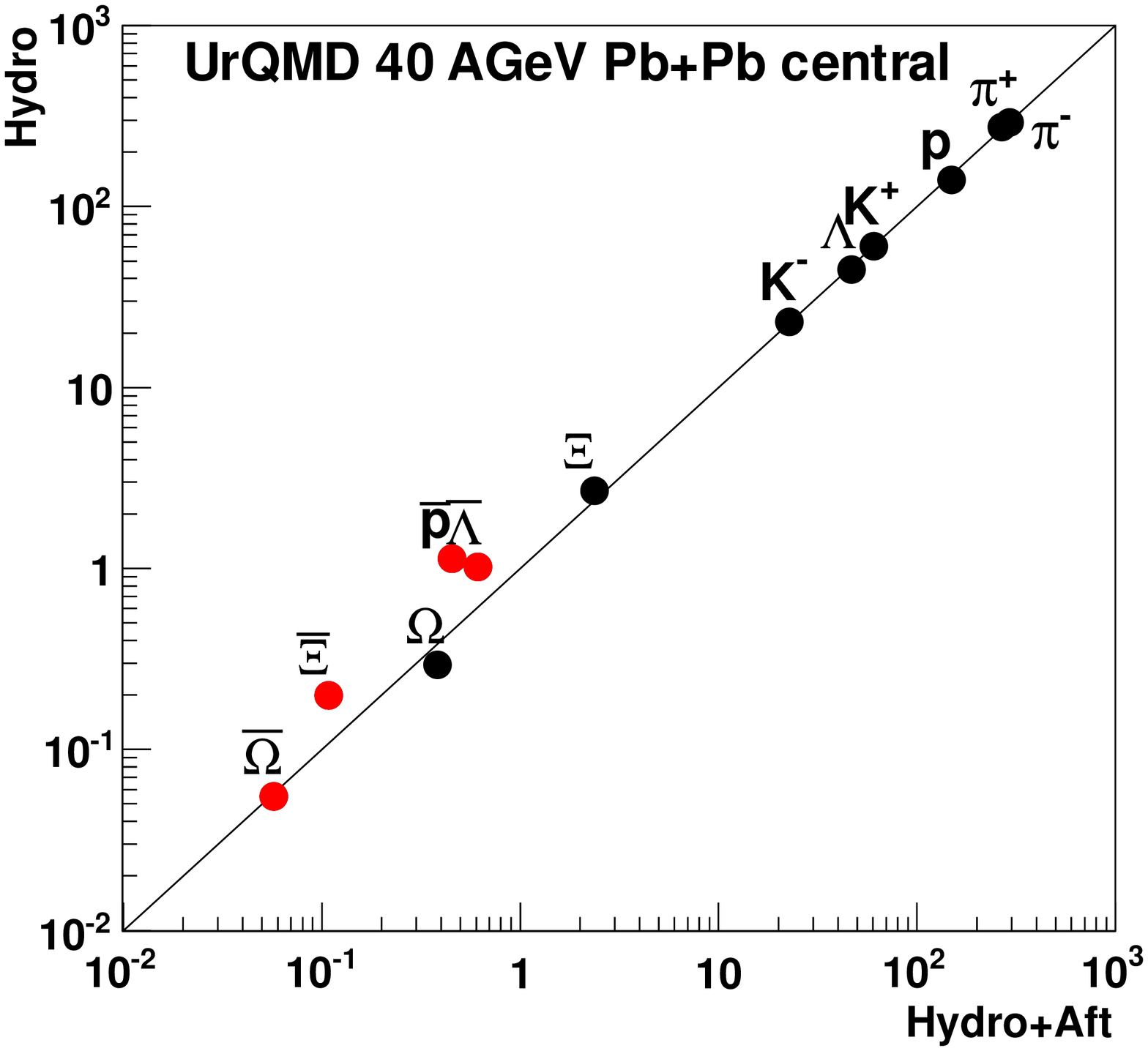}
\includegraphics[width=0.4 \textwidth]{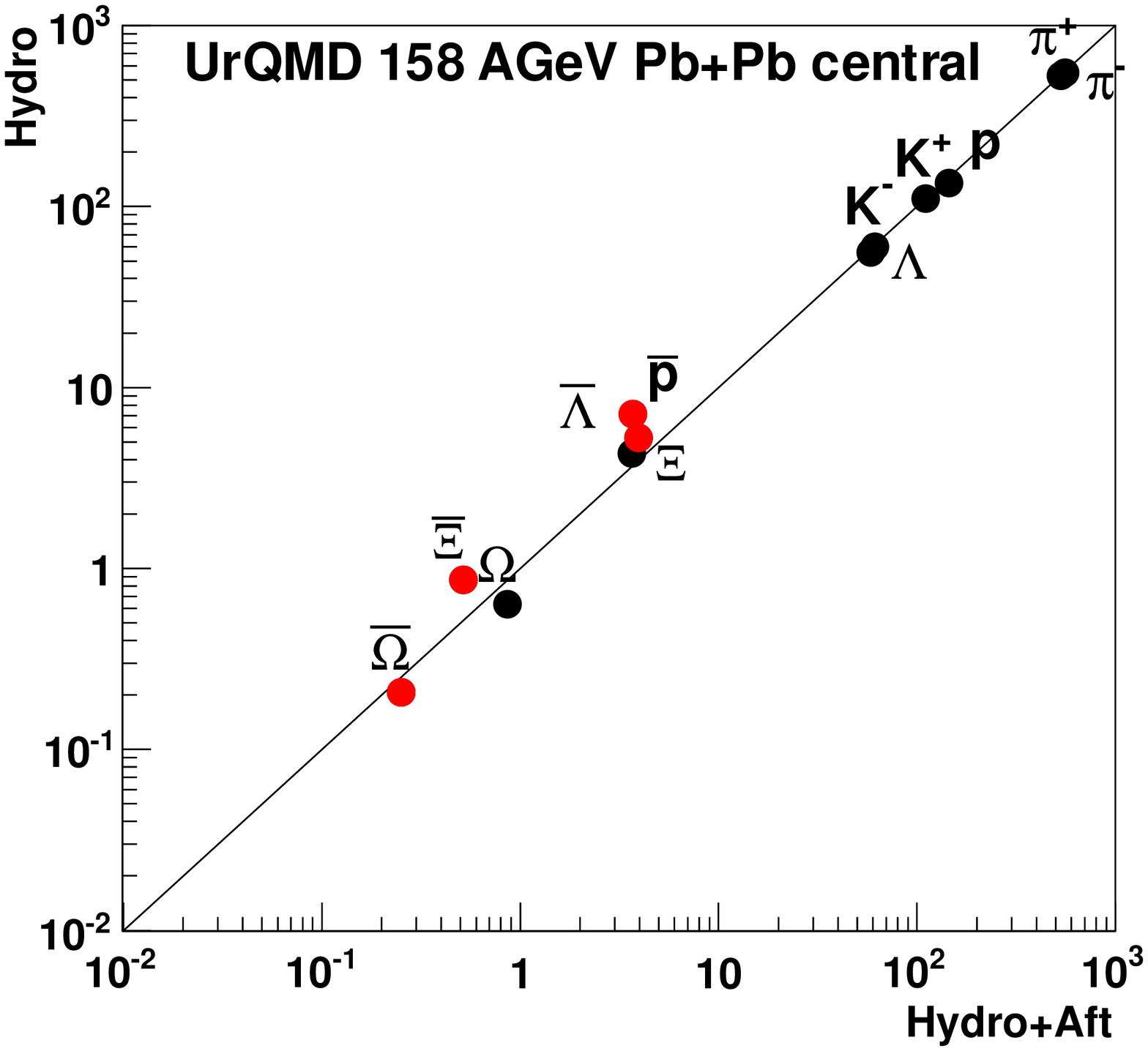}
\end{center}
\caption{UrQMD calculations for central Pb+Pb at 40$A$ and 158$A$~GeV, plotting the result before the hadronic cascade vs.\ the result after.}
\label{fig:2}
\end{figure}

Figure~\ref{fig:2} shows the effect of the final UrQMD cascade stage, in a plot of hadron multiplicities directly after the hydro stage, vs. the multiplicities at the end of the cascade. We illustrate these conditions for central Pb+Pb collisions at the SPS energies $40A$ and $158A$~GeV. We see the bulk hadrons unaffected by the afterburner, including the $\Xi$, $\Omega$ and $\bar{\Omega}$. Whereas the other antibaryons, $\bar{\mathrm{p}}$, $\bar{\Lambda}$ and $\bar{\Xi}$, are significantly and selectively suppressed.

\begin{figure}
\begin{center}
\includegraphics[width=0.4 \textwidth]{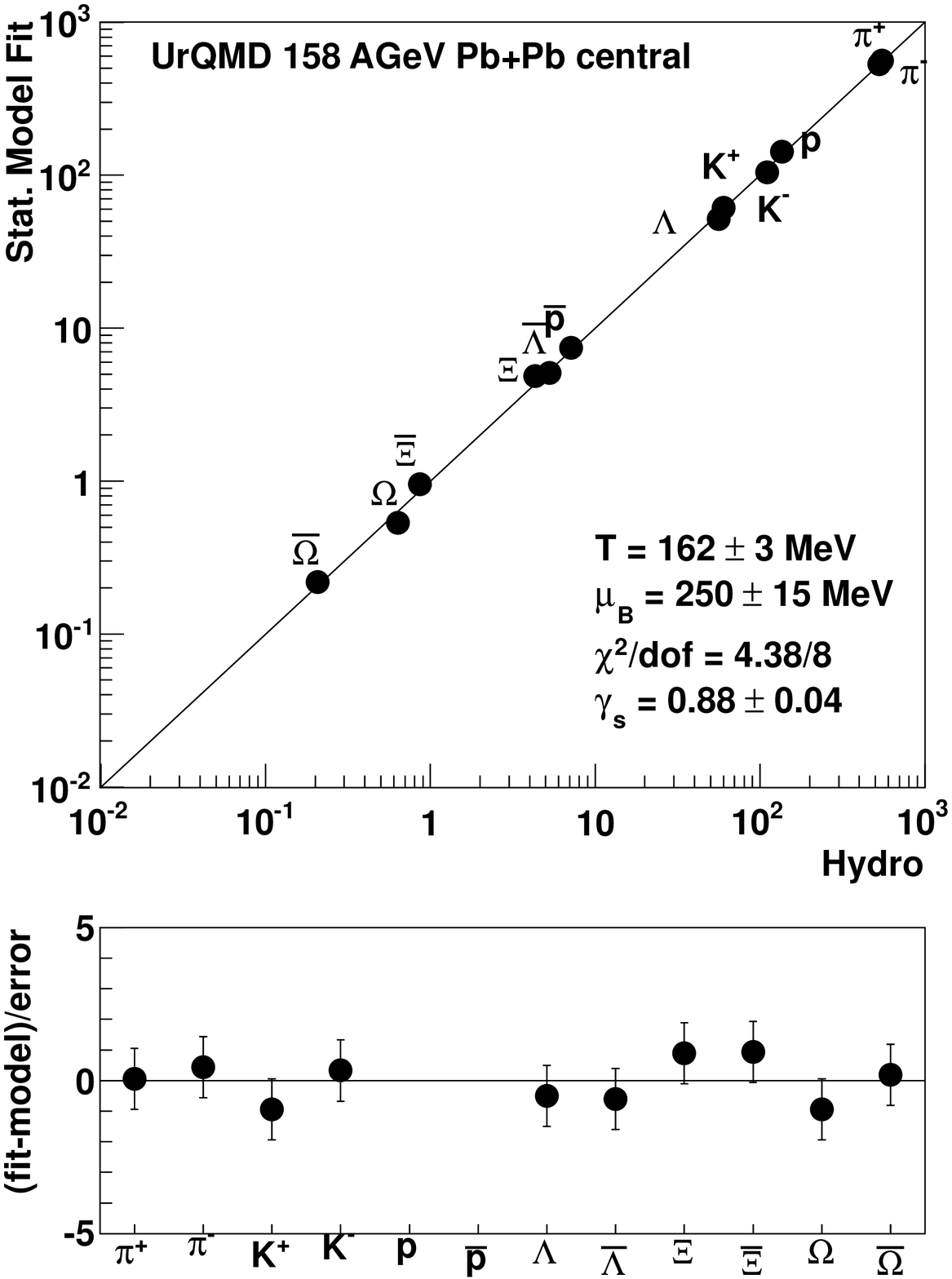}
\includegraphics[width=0.4 \textwidth]{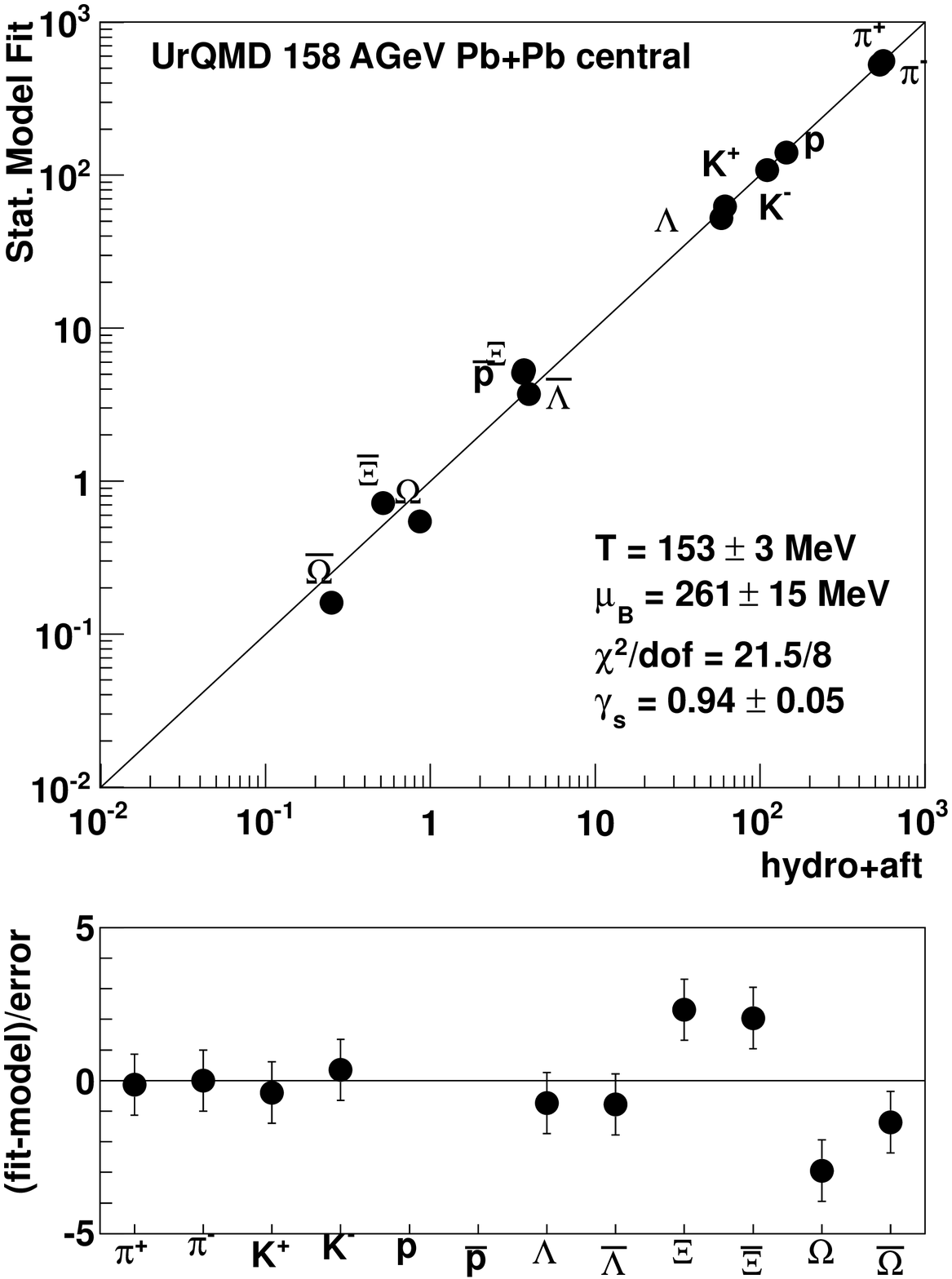}
\end{center}
\caption{Statistical model fit to UrQMD results for central Pb+Pb at 158$A$~GeV, before and after the cascade stage.}
\label{fig:3}
\end{figure}

\begin{figure}
\begin{center}
\includegraphics[width=0.4 \textwidth]{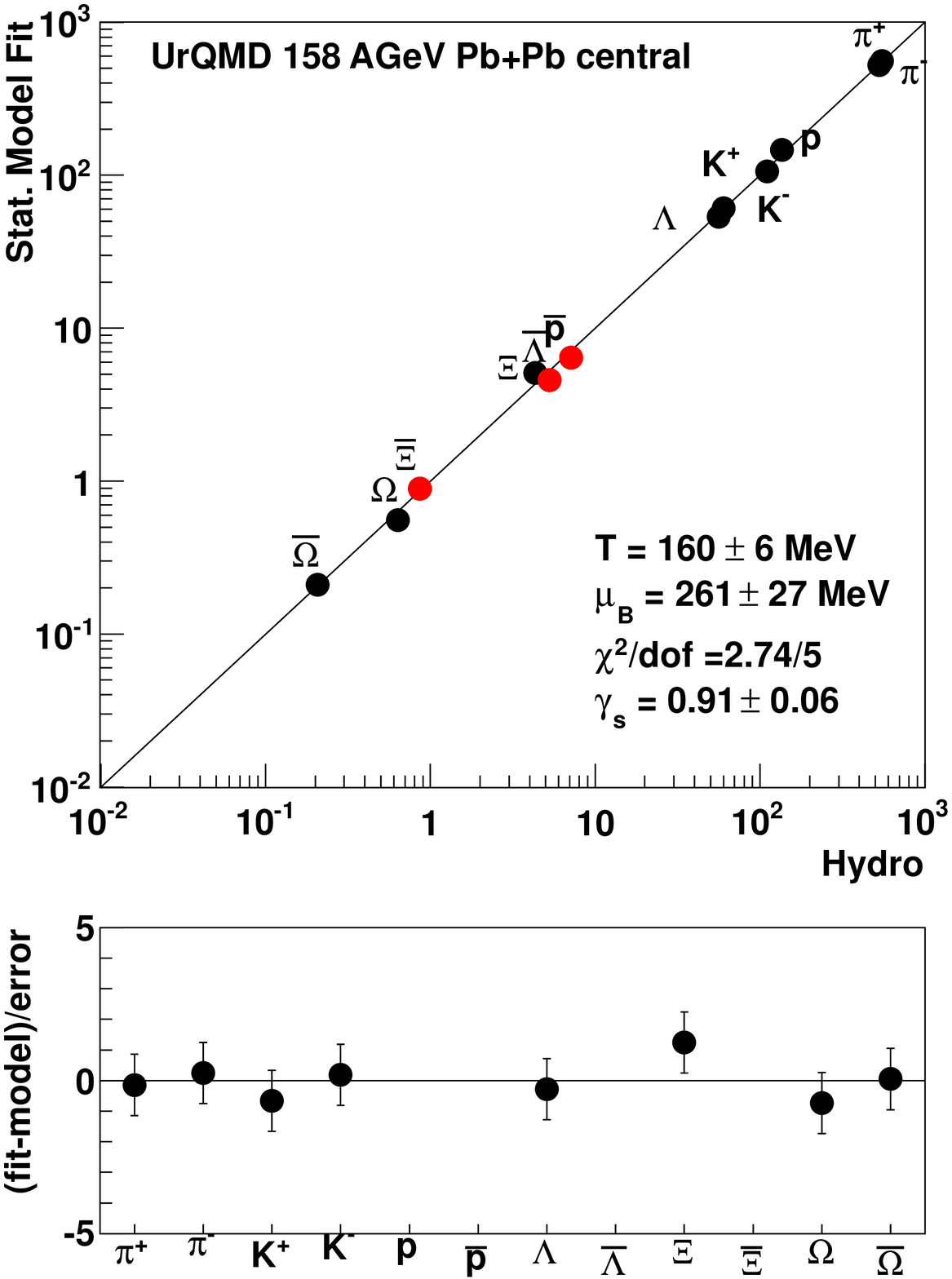}
\includegraphics[width=0.4 \textwidth]{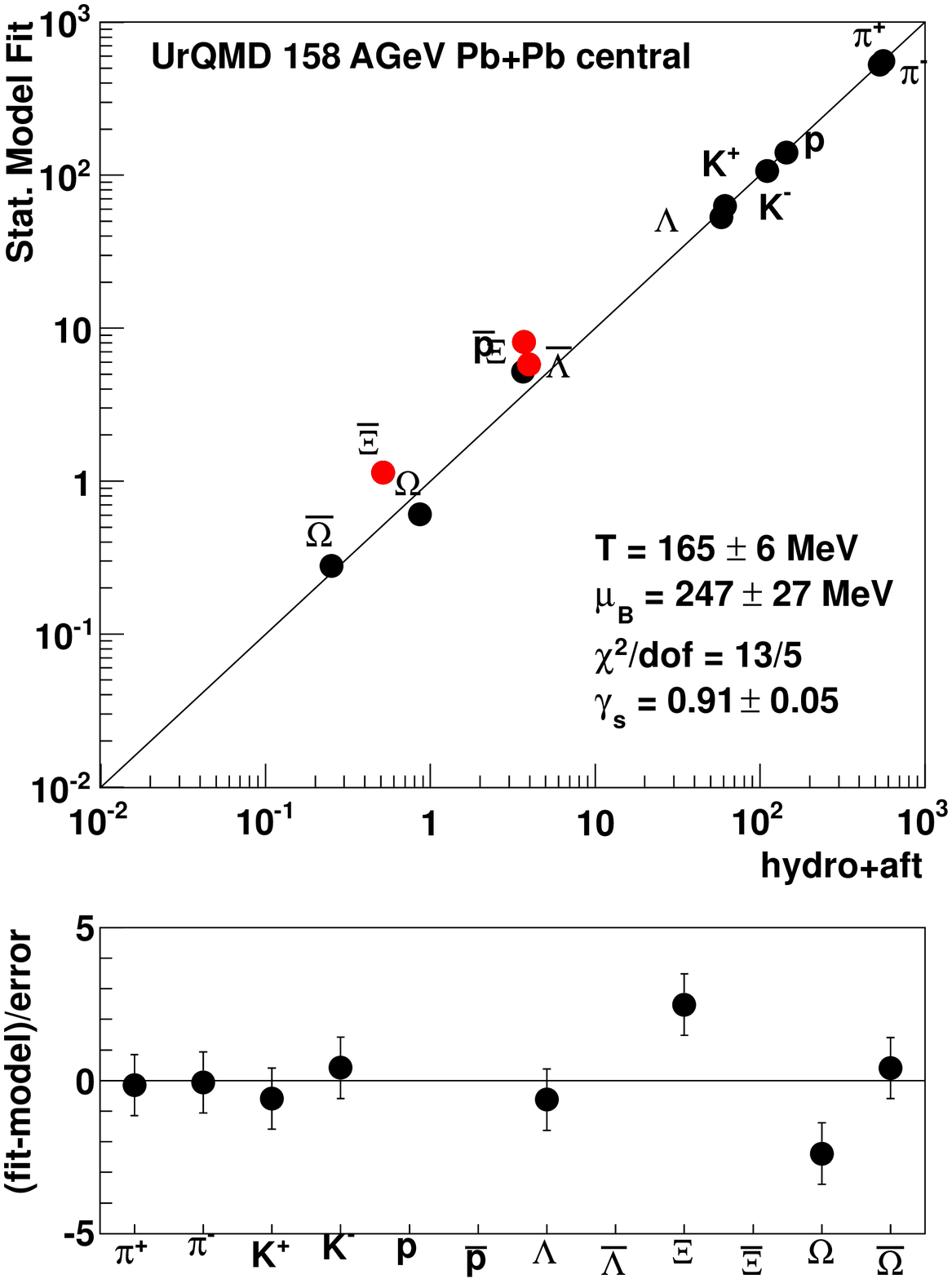}
\end{center}
\caption{The same as figure~3 except that antibaryons are excluded from the SM fit procedure (but included in the figure).}
\label{fig:4}
\end{figure}

Figure~\ref{fig:3} illustrates the fits to the hybrid UrQMD results by the statistical model, choosing the $158A$~GeV case as an example. The afterburner stage indeed shifts ($T, \mub$) considerably, from (162, 250) to (153, 261). However, note the dramatic decrease of fit quality, from 4.4 to 21.5 in $\chi^2$. The effect of the afterburner is, thus, not an in-equilibrium cooling but rather a distortion of the hadron yield distribution, away from equilibrium - as we could guess from figure~\ref{fig:2} already. The idea arises to exclude $\bar{\mathrm{p}}$, $\bar{\Lambda}$ and $\bar{\Xi}$ from the SM fit. Figure~\ref{fig:4} shows an example, again at $158A$~GeV. No cooling occurs. The fit to the afterburner output (which features a tolerable $\chi^2$) now ignores the far off-diagonal $\bar{\mathrm{p}}$, $\bar{\Lambda}$ and $\bar{\Xi}$ entries.

We conclude that the hadron/resonance cascade as modelled in the microscopic dynamics of UrQMD can NOT transport an initially established hadrochemical equilibrium from the phase coexistence line of figure~\ref{fig:1}, down to the freeze-out line. However, it distorts the hadron yield distribution
which leads to a downward shift of the freezeout parameters derived from SM analysis, albeit at the cost of rather unsatisfactory $\chi^2$.

\begin{figure}
\begin{center}
\includegraphics[width=0.4 \textwidth]{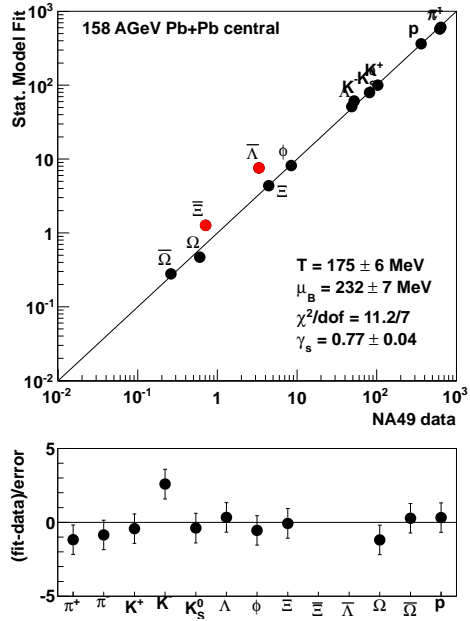}
\end{center}
\caption{Statistical Model fit to the NA49 data~\cite{10} for central Pb+Pb at 158$A$~GeV, omitting the antibaryons except anti-$\Omega$ from the fit procedure (their corresponding yields are shown).}
\label{fig:5}
\end{figure}

Turning to SM analysis of real SPS data, we note, first, that the $\chi^2$ values obtained from a parallel SM analysis of the NA49 data~\cite{10}, which are exhibited in figure~\ref{fig:1}, were also found to be rather high. The idea thus arises to suppose effects in the data, similar to some extent to our UrQMD findings. Figure~\ref{fig:5} thus shows a prediction of the SM to the NA49 data at $158A$~GeV where $\bar{\mathrm{p}}$, $\bar{\Lambda}$ and $\bar{\Xi}$ are excluded from the fit. Very much reminiscent of figure~\ref{fig:4} (right), the SM fit here moves up from $T=158$~MeV (figure~\ref{fig:1}) to above 170~MeV, at reasonable $\chi^2$: far above the conventional freeze-out curve at this energy. We shall thus revisit the freeze-out curve at \mub\ from about 280 to 430~MeV, obtained, hitherto, from an ``unfiltered'' application of the statistical model. At the much higher RHIC energies, with \mub\ approaching zero, we may expect
a smaller such effect of the cascade stage because of the approximate baryon-antibaryon symmetry.

\section*{References}


\begin{thebibliography}{99}

\bibitem{Montvay:1978sv}
  I.~Montvay, J.~Zimanyi,
  Nucl.\ Phys.\  {\bf A316}, 490 (1979).

\bibitem{Stock:1985xe}
  R.~Stock,
  Phys.\ Rept.\  {\bf 135}, 259-315 (1986).

\bibitem{Becattini:2003wp}
  F.~Becattini, M.~Gazdzicki, A.~Keranen, J.~Manninen, R.~Stock,
  Phys.\ Rev.\  {\bf C69}, 024905 (2004).
  [hep-ph/0310049].

\bibitem{PBM_QGP3}
P.~Braun-Munzinger, K.~Redlich and J.~Stachel,
in Quark-Gluon Plasma 3, eds. R.~C.~Hwa and X.~N.~Wang, World Scientific 2004, p. 491.


\bibitem{3}
J.~Cleymans, H.~Oeschler, K.~Redlich and S.~Wheaton, Phys.\ Rev.\ {\bf C73}, 034905 (2006).

\bibitem{4}
F.~Karsch, PoS(Lattice07) 015.
 
\bibitem{5}
R.~Rapp, T.~Schaefer and E.~V.~Shuryak, Ann.\ Phys.\ 280, 35 (2000).

\bibitem{6}
P.~Braun-Munzinger, J.~Stachel and Ch.~Wetterich, Phys.\ Lett.\ B {\bf 596}, 61 (2004), \\
R.~Stock, [arXiv:0703050, nucl-th].

\bibitem{7}
L.~McLerran and R.~Pisarski, Nucl.\ Phys.\ A {\bf 796} (2007), 83, \\
L.~McLerran, [arXiv:0812.1518 [hep-ph]].

\bibitem{8}
  H.~Petersen, J.~Steinheimer, G.~Burau, M.~Bleicher, H.~Stocker,
  Phys.\ Rev.\  {\bf C78}, 044901 (2008).
  [arXiv:0806.1695 [nucl-th]].

\bibitem{9}
F.~Becattini, J.~Manninen and M.~Gazdzicki,
 Phys.\ Rev.\  C {\bf 73} (2006) 044905
 [arXiv:hep-ph/0511092].

\bibitem{10}
\texttt{https://edms.cern.ch/file/1075059/1/na49\_compil.pdf}
and references therein.

\end{thebibliography}
\end{document}